\documentclass[aip,reprint]{revtex4-1} 
\usepackage{graphicx}
\usepackage{amsmath}
\usepackage[dvipsnames]{xcolor}
\usepackage[colorlinks=true, allcolors=blue]{hyperref}
\usepackage[a4paper, total={6in, 8in}]{geometry}
\usepackage{float}
\usepackage[parfill]{parskip}
\usepackage{dcolumn}
\usepackage{relsize}

\begin{document}

\title{Electrical coupling of superparamagnetic tunnel junctions mediated by spin-transfer-torques}
\author{Leo Schnitzspan}
\author{Mathias Kläui}
\author{Gerhard Jakob}
\affiliation{Institute of Physics, Johannes Gutenberg-University Mainz, 55122 Mainz, Germany}
\affiliation{Max Planck Graduate Center Mainz, 55122 Mainz, Germany}

\date{\today}

\begin{abstract}
In this work, the effect of electrical coupling on stochastic switching of two in-plane superparamagnetic tunnel junctions (SMTJs) is studied, using experimental measurements as well as simulations.
The coupling mechanism relies on the spin-transfer-torque (STT) effect, which enables the manipulation of the state probability of an SMTJ.
Through the investigation of time-lagged cross-correlation, the strength and direction of the coupling are determined.
In particular, the characteristic state probability transfer curve of each SMTJ leads to the emergence of a similarity or dissimilarity effect.
The cross-correlation as a function of applied source voltage reveals that the strongest coupling occurs for high positive voltages for our SMTJs.
In addition, we show state tuneability as well as coupling control by the applied voltage.
The experimental findings of the cross-correlation are in agreement with our simulation results.
\end{abstract}

\maketitle


Magnetic tunnel junctions (MTJs) consist of two ferromagnetic layers, separated by an insulating layer, and exhibit a large resistance change from the parallel to the antiparallel state, caused by the tunnel magnetoresistance effect (TMR).
Due to low energy consumption for the non-volatile configuration and CMOS compatibility, MTJs are well suited for memory devices such as random access memory (MRAM) \cite{MRAM_overview1}.
However, if the energy barrier between the parallel (P) and antiparallel (AP) state is small, thermal excitation can induce spontaneous fluctuations between these states \cite{Fukami_theory}.
In recent studies, random fluctuations in superparamagnetic tunnel junctions (SMTJs) have been demonstrated across time scales ranging from milliseconds to nanoseconds \cite{Reiss_smtj, Majetich_smtj, ip_SMTJ_ns_Sun, ip_SMTJ_ns_Fukami, LS_SMTJ}.
Moreover, the potential of SMTJs for generating true random numbers has been recognized \cite{LS_SMTJ, RNG1, RNG2, RNG3}, which holds particular significance for cryptographic applications that require a high quality of randomness.
Additionally, SMTJs are promising candidates in neuromorphic spintronics, as artificial synapses or neurons \cite{neuro_spintronics_rev1}.
In larger spintronic systems coupling between individual elements can occur due to spin torques acting on the magnetization of these devices.
For instance, previous investigations demonstrated the electrical coupling of spin-torque nano oscillators (STNO) \cite{SHNO1}.
Understanding and investigating the coupling behavior of stochastic MTJs is of paramount importance for successful implementations of probabilistic systems, such as Boltzmann machines \cite{BM_Camsari}.
The coupling between SMTJs refers to the interaction and influence of one junction on another, which can be established through various effects, including dipole coupling \cite{SMTJ_dipole_coupling1}, strain \cite{SMTJ_ensemble_strain}, or electrical interaction via spin-transfer-torques (STT) \cite{el_coupled_MTJs, el_coupled_SMTJs2}.

In this work, we explore the effects of coupling on the stochastic switching via electrical interconnection, which offers advantages in terms of scalability, rapid mediation of coupling, and ease of device implementation.
So far electrical coupling of two-level superparamagnetic tunnel junctions in parallel circuitry has only been studied under the influence of external magnetic fields \cite{el_coupled_MTJs} or via an applied current source \cite{el_coupled_SMTJs2}.
However, for practical applications, these approaches might not be ideal in terms of feasibility and therefore we focus on the tuneability of two SMTJs in series under an applied DC voltage.
Electrical coupling might also be promising for probabilistic networks based on stochastic MTJs, such as Boltzmann \cite{BM_Camsari} or Ising machines \cite{Ising_machine_Camsari}.


TMR stacks were deposited at room temperature on oxidized Si substrates using rf- and dc-magnetron sputtering (Singulus Rotaris) with the following composition (film thickness in nanometer): Ta(10)/Ru(10)/Ta(10)/PtMn(20)/CoFe(2.2)/
Ru(0.8)/ CoFeB(2.4)/MgO(1.1)/CoFeB(3.0)/Ta
(10)/Ru(20) and is based on an optimized stack, developed earlier \cite{LS_ADI}.
The TMR ratio of our stack is found to be approximately $100\,\%$ and the resistance area (RA) product around $15\,\Omega$µm$^2$.
Annealing was carried out in a $300\,$mT in-plane field at $300\,^{\circ}$C for $1\,$h.
An in-plane field of a few mT is typically applied in order to compensate for stray fields from other ferromagnetic layers in the stack and to tune the stochastic switching.
Nanopillars were patterned in circular shapes of diameters of approximately $60\,$nm.

\begin{figure*}
\centering
\includegraphics[width=0.98\textwidth]{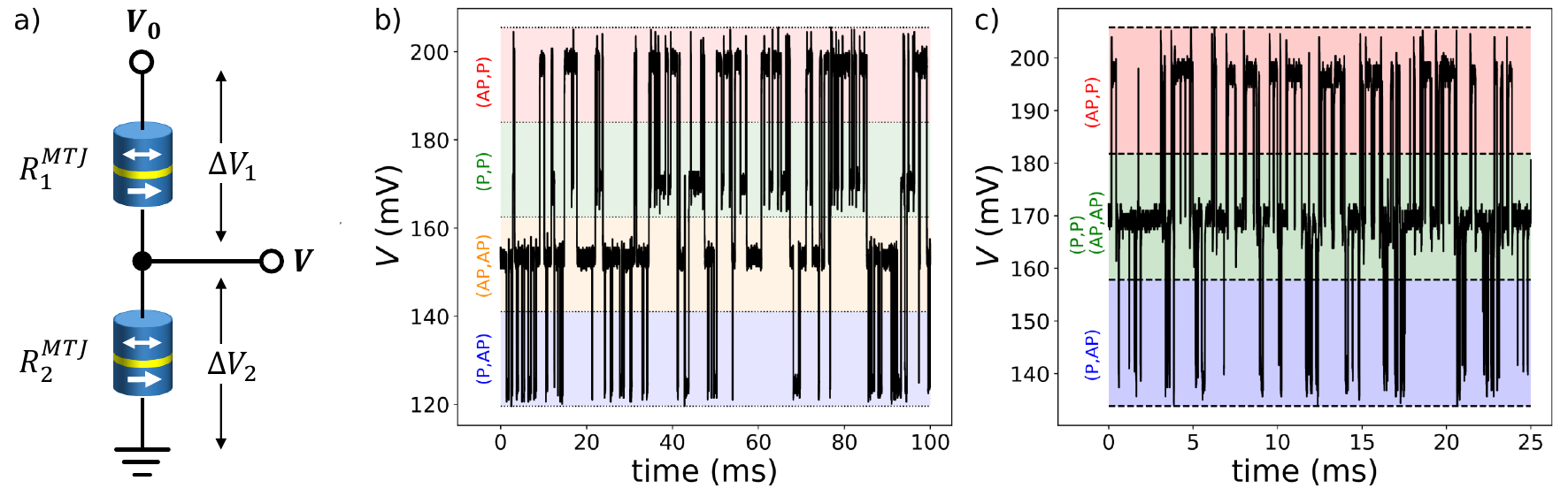}
\caption{\label{fig:setup}a) Sketch of the measurement set-up of two SMTJs in series. The source voltage $V_0$ is applied and the voltage drop $V$ between both SMTJs is measured. A voltage difference across an MTJ resistance $R_i^{MTJ}$ is indicated by $\Delta V_i$. b), c) Oscilloscope measurements of two coupled SMTJs ($V_0=0.3\,$V). The resistance band of three or four separate states (P,P), (AP,P), (P,AP) and (AP,AP) is depicted by background colors.}  
\end{figure*}

After nanopillar patterning, two stochastic MTJs were interconnected in series, as illustrated in Figure \ref{fig:setup}a.
The MTJ resistance states are in the k$\Omega$ range and the resulting volatile voltage drop, as shown in Fig. \ref{fig:setup}a and \ref{fig:setup}b, is acquired by an oscilloscope (Tektronix DPO7543).
Depending on the resistance states of both SMTJs, either three or four states are effectively distinguishable, since two different states can result in almost the same voltage drop $R_B / ( R_A + R_B)$.
In particular, the combination of SMTJ $B$ and $C$ provides similar voltage outputs for the states (P,P) and (AP,AP), leading to a total of three states, as shown in Fig. \ref{fig:setup}c.
To describe the coupled mixed states of two SMTJs $A$ and $B$ we use a bracket notation: (S$_A$,S$_B$), 
where $S_i$ stands for the current state (P or AP) of the $i^{th}$ MTJ.
%


We conducted simulations to investigate the stochastic fluctuations of two electrically coupled superparamagnetic magnetic tunnel junctions arranged in series.
We consider that each SMTJ on its own behaves like a Bernoulli distributed random variable of probability $p$.
For a specific voltage and temperature, an SMTJ $i$ has a certain probability $p_i$ to occupy the AP state.
A modification in the bias voltage across the junction will result in an altered probability $p_i$ due to the STT effect.
The relation between AP probability and the applied voltage is here called PV-transfer function.
However, the (parallel or antiparallel) resistance also exhibits non-linear behavior in response to the bias voltage, which is denoted as the RV-transfer function and must be considered in the simulation.
Due to the superparamagnetic property of the MTJ free layer, random resistance fluctuations (also called telegraph noise) occur, which can be characterized by dwell times $\tau$ of the P- and AP-state. 
The dwell time in the macrospin approximation for $E_b/k_{\mathrm{B}} T \geq 1$, follows the Néel-Arrhenius law \cite{theory_Neel_law}:
$\tau = \tau_0 \, \exp \left( E_b/k_{\mathrm{B}} T \right)$,
where $E_b$ is the energy barrier between the states, $k_{\mathrm{B}}$ the Boltzmann constant, $T$ the temperature and $\tau_0$ the attempt time.
The assumption of the Néel-Arrhenius model is switching between two distinct energy minima, i.e.\ two distinct resistance states.
For simplicity, we use $\Delta=E_b/k_{\mathrm{B}} T$ in the following.
Dwell times can be manipulated by applying a current or voltage to the tunnel junction, which affects the effective energy barrier through the effect of spin-transfer torque \cite{STT_theory1, STT_theory2}:
\begin{equation}
\label{eq:neel_arrhenius}
    \tau_{p,ap} = \tau_0 e^{ \Delta (1 \pm V/V_c)}
\end{equation}
Here, $\tau_0$ is the attempt time, $\Delta$ is the (unitless) energy barrier, $V$ is the voltage across the junctions, and $V_c$ is the critical voltage for deterministic switching at $0\,$K \cite{STT_theory1, STT_theory2}.
The dwell time ratio $\tau_{ap}/(\tau_{ap} + \tau_{p})$ then describes the probability to observe the AP state.
With equation \ref{eq:neel_arrhenius} we can derive the characteristic voltage-dependent AP probability. 
\begin{align}
\label{eq:PV_sigmoid}
    P_{ap}(V) &= \frac{\tau_{ap}}{\tau_{ap} + \tau_{p}} = \nonumber
    \frac{e^{ \Delta (1-V/V_c)}}{e^{ \Delta (1-V/V_c)} + e^{ \Delta (1+V/V_c)}} \\ 
    &= \frac{1}{e^{ 2\Delta V/V_c} + 1} 
\end{align}
Here, $\Delta$ is the energy barrier, $V$ the applied voltage and $V_c$ the critical voltage.
Equation \ref{eq:PV_sigmoid} represents the typical sigmoid relation and is used to model the PV-transfer function for both SMTJs.

For the generation of artificial random telegraph noise, which can be considered as a two-state Markov process, the Poisson distribution of dwell times has to be considered.
Telegraph noise is observed in various physical systems, like flash memory \cite{RTN_flash} or field effect transistors \cite{RTN_MOSFET}, and is characterized by an exponential distribution of dwell times.
Hence, the Poisson process can describe the probability of a stochastic MTJ switching to another state \cite{Fukami_theory}.
Moreover, at low current densities in stochastic MTJs, the switching event follows a probability density function (PDF) given by \cite{MTJ_switching1}:
$f^{sw}(t) = 1/ \tau \cdot e^{-t/\tau}$ for a mean dwell time $\tau$, which is the inverse of the average switching rate.
Therefore, the thermal switching probability that the magnetization switches within time $t$ in an MTJ with modified energy barrier can be expressed as \cite{STT_theory1}: 
\begin{align}
\label{eq:P_switch}
    F^{sw}(t) &= 1 - \exp \left(-t / \tau_0 e^{-\Delta (1\pm V/V_c)} \right) \nonumber \\
    &= 1 - e^{-t / \tau} \\ \nonumber
    &\approx t/\tau \qquad \qquad (t\ll \tau_{ap}) \\ \nonumber
\end{align}
which is the cumulative distribution function (CDF) of $f_{sw}(t)$ \cite{MTJ_switching2}.
As a consequence, the switching probabilities from the P to the AP state and vice versa can be expressed as $P^{sw}_{p \rightarrow ap} = 1 - e^{-t / \tau_p}$, $P^{sw}_{ap \rightarrow p} = 1 - e^{-t / \tau_{ap}}$.
These two switching probabilities are relevant for a single stochastic MTJ.
However, for two coupled SMTJs there are 12 different switching probabilities, due to the electrical coupling.
In general, for $n$ SMTJs in series there exist $2^n$ distinct Markov states with $4^n-2^n$ 
transition probabilities.

\begin{figure*}[t]
\centering
\includegraphics[width=0.99\textwidth]{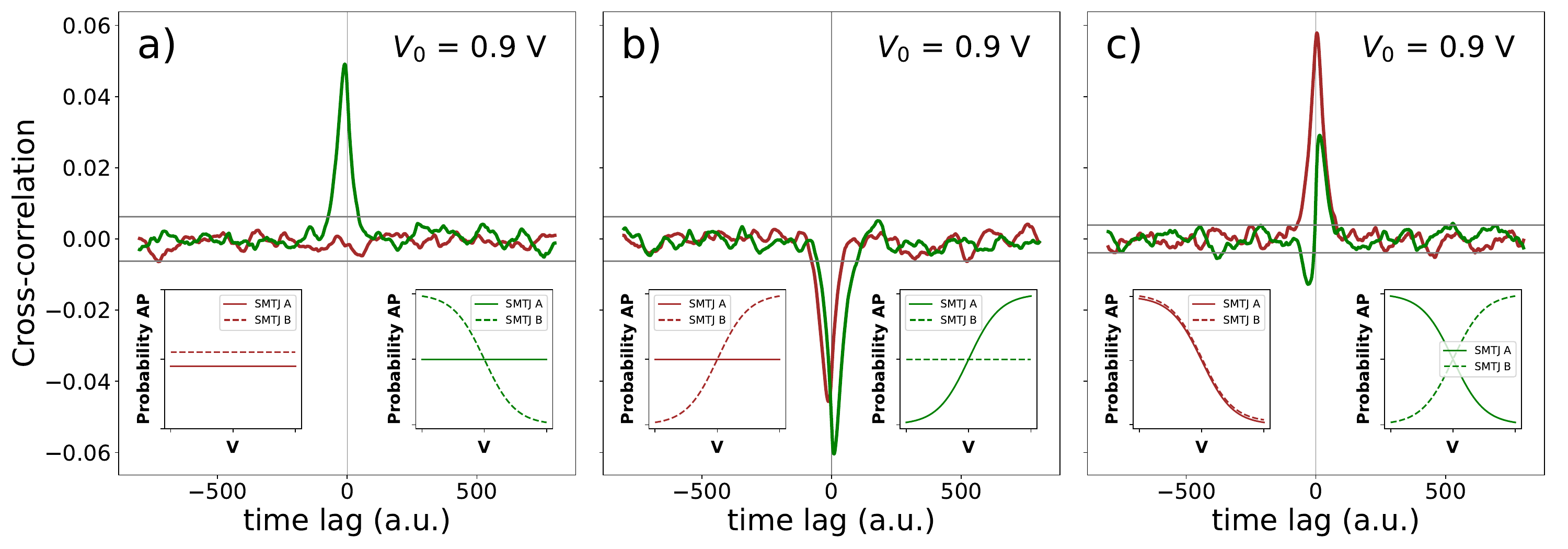}
\caption{\label{fig:cross_corr_peaks}A simulation of the cross-correlation depending on different PV-transfer curves. a) In red the cross-correlation function is shown for two constant PV-transfer curves, as depicted by the left inset. In green the correlation is shown for a sigmoid transfer curve for SMTJ $A$ and a constant transfer curve for SMTJ $B$, as depicted by the right inset. b) and c) demonstrate the effect on cross-correlation for different PV-transfer curves of SMTJ $A$ and $B$.}        
\end{figure*}

To simulate a time series of these Markov states for two coupled SMTJs, we start by choosing average dwell times $\tau = \sqrt{\tau_p \tau_{ap}}$ for both SMTJs and a time step $dt$ for each iteration.
During the time step $dt$ the switching probability $P^{sw}$ for each SMTJ can be calculated according to equation \ref{eq:P_switch} (for $dt \rightarrow 0$ also $P^{sw} \rightarrow 0$).
Next, we draw a binary random variable from a Bernoulli distribution: $\boldsymbol{\mathcal{B}}(P_i^{sw})$ for SMTJ $i$.
If switching occurs ($S_i \rightarrow \overline{S_i}$) then this will affect the switching probability $P_i^{sw}$ of both SMTJs, due to a change in the voltage drop $\Delta V_i$ caused by a change in the resistance states $R_i$.
Since for each time step the MTJ states are known, the resistance states can be derived.
However, because the MTJ resistance depends on the voltage (RV-transfer function) due to the tunneling, the MTJ resistance and voltage drop will both affect each other until an equilibrium is found.
This problem is solved by a recursive function of $V$ and $R$ in the simulation.
The (equilibrium) voltage drop $\Delta V_i$ is then used to calculate the new switching probabilities $P_i^{sw}$ for each SMTJ, according to the characteristic PV-transfer functions.
From the new switching probabilities, new MTJ states can be sampled and the simulation process starts from the beginning, which ensures a sequential series of Markov states.
In order to obtain good statistical results, stochastic time series were simulated for $10^7$ time steps (iterations).
For the simulation the PV-transfer curves are modeled by a sigmoidal function (see equation \ref{eq:PV_sigmoid}), whereas
the RV-transfer curves are described by an approximation according to Brinkman \cite{tunnel_current_brinkman}.
The reduction in resistance for higher absolute bias voltages is due to the non-linear tunneling current and can be represented by the following function \cite{tunnel_current_brinkman}:
$R(V)=a/(b + c V^2) + d$, where $a, b, c$ and $d$ are (fitting) parameters.

In order to evaluate the coupling strength, a cross-correlation function of simulated (or measured) time series data can be determined. 
Cross-correlation is commonly used in signal processing for pattern recognition \cite{XC_pattern_recognition} or as a measure of similarity of two series ($A$ and $B$) as a function of the relative shift of one series with respect to the other.
In our case, the cross-correlation of two SMTJ time series is calculated.
A positive (negative) time lag corresponds to a positive (negative) shift of time series $B$ with respect to $A$.
The cross-correlation function is defined as:
\begin{equation}
\label{eq:xcorr}
    C(t) = \frac{ \sum_i \frac{1}{N}(A_i - \overline{A})  (B_{i-t} - \overline{B})}{s_A s_B}
\end{equation}
where $t$ is the (time) lag ($t \in [-N/2, ..., N/2]$ in multiples of the incremental time step $dt$), $\overline{A}$, $\overline{B}$ the mean and $s_A$, $s_B$ the standard deviation of time series $A$ and $B$ with (time) length $N$.
The values $A_i$ and $B_i$ with $i \notin [0, 1, ..., N-1]$ are 0 and do not contribute to the cross-correlation.\\

\begin{figure*}[t]
\centering
\includegraphics[width=0.99\textwidth]{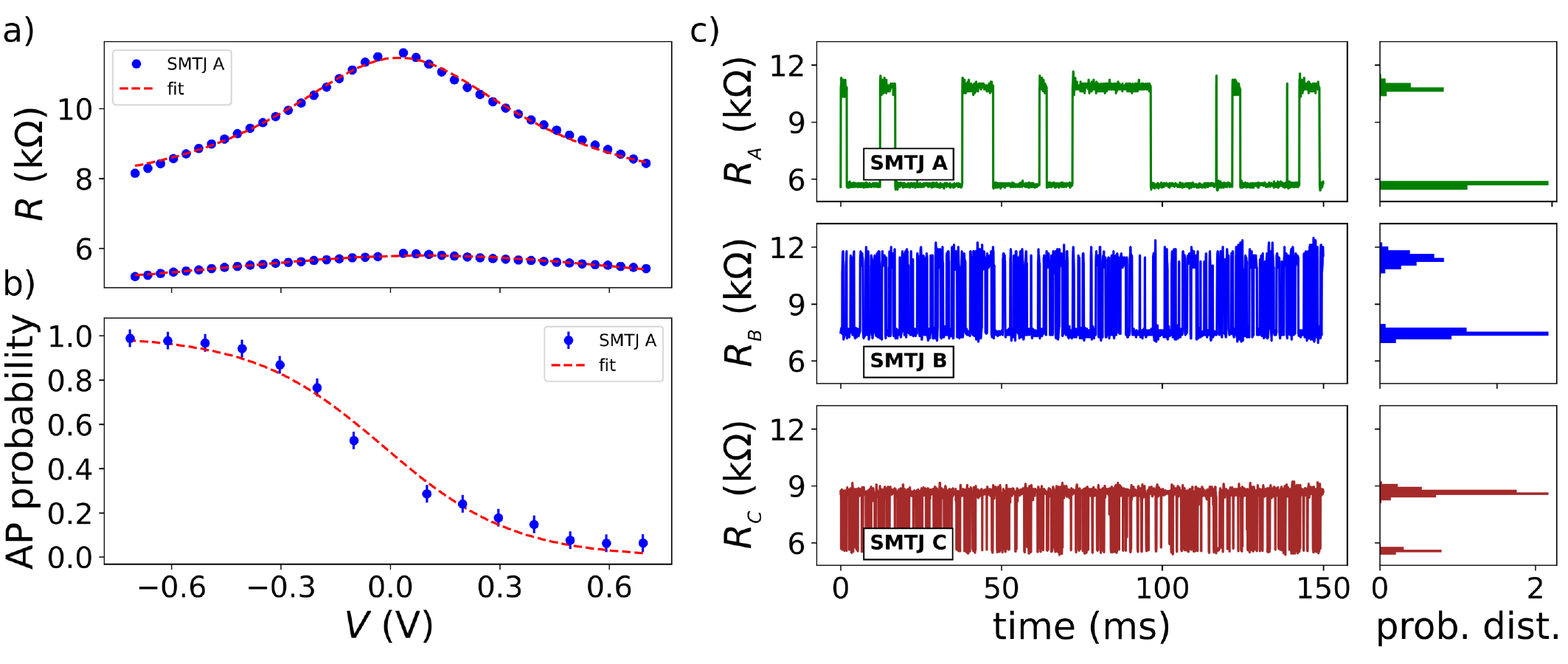}
\caption{\label{fig:SMTJ_characterization}a) MTJ resistance as a function of applied bias voltage for a fixed free layer. b) AP probability transfer curve for the stochastic MTJ with sigmoid fit in red. c) Time series of three different SMTJs $A$, $B$ and $C$ along with histograms illustrating their respective probability distributions.}         
\end{figure*}
%
%
\begin{table}[b]
\caption{\label{tab:cross_corr}Cross-correlation summary}
\begin{center}
\begin{tabular}{c c c c}
     Lag & Correlation & Direction & Effect \\ [0.5ex] 
     \hline\hline
     + & + & $B \rightarrow A$ & similarity \\ 
     \hline
     + & - & $B \rightarrow A$ & dissimilarity  \\
     \hline
     - & + & $A \rightarrow B$ & similarity  \\
     \hline
     - & - & $A \rightarrow B$ & dissimilarity  \\
     \hline
\end{tabular}
\end{center}
\end{table}
Figure \ref{fig:cross_corr_peaks} highlights the simulation results, assuming instantaneous state transitions, the same average switching rates, and the absence of additional noise.
The cross-correlation function reveals a peak at 0 time lag, which is expected in the context of two interacting time series.
This behavior arises due to the influence of switching events in one MTJ on the state of the other MTJ, caused by changes in STT.
The correlation amplitude primarily depends on the TMR ratios and the applied voltage to the coupled SMTJs, since the SMTJs are most "sensitive" to voltage drops in regions where the gradient of the PV-transfer function is maximal.
The correlation peaks appear to be positive or negative and to be shifted to positive or negative time lags.
If a peak is on the positive side of the cross-correlation function, then the effect of series $B$ on $A$ is obtained, while for the negative side, the effect of $A$ on $B$ is accessed.
Therefore, we are able to measure the effect of both SMTJs on each other.
A positive correlation would coincide with a preference of both SMTJs to stay in the same state ("similarity effect"), while a negative correlation coincides with a preference for antiparallel alignment ("dissimilarity effect").
These findings are illustrated in Fig. \ref{fig:cross_corr_peaks} and summarized in Table \ref{tab:cross_corr}.
In Fig. \ref{fig:cross_corr_peaks}a the cross-correlation is simulated for two SMTJs, where the coupling is artificially turned off, (red curve) which corresponds to constant PV-curves, and for the case where only the PV-curve of SMTJ $B$ is constant, while the one for SMTJ $A$ shows the sigmoid function (green curve).
In the first case, no cross-correlation peak is observed, while for the second case, a positive peak emerges in the negative time lag range, which indicates that SMTJ $A$ affects SMTJ $B$ and $B$ prefers to adopt to the same state than $A$.
Fig. \ref{fig:cross_corr_peaks}b illustrates the effect of switching the PV-transfer curves of $A$ and $B$, as shown in the insets, where as a consequence also the correlation peak shifts from negative (red curve) to positive (green curve) time lag range.
A more realistic simulation is depicted in Figure \ref{fig:cross_corr_peaks}c for two coupled SMTJs with sigmoid PV-transfer functions and bilateral coupling.
If both PV-transfer functions exhibit the same trend, such as both monotonically increasing, it will lead to one symmetric correlation peak centered at 0.
However, if one of the PV-transfer functions is reversed ($f(x) \rightarrow f(-x)$), it results in an anti-symmetrical peak, with positive as well as negative correlations close to 0 time lag, each peak respectively corresponding to an SMTJ.
%

In order to verify our simulation results, we conducted time series measurements for different combinations of SMTJs, as illustrated in Figure \ref{fig:setup}.
Depending on the resistance states of each MTJ, this results in four possible resistance states leading to four different voltage drops.
Figure \ref{fig:SMTJ_characterization}c displays the characteristic time series of SMTJs $A$, $B$ and $C$ for $0.2\,$V.
Due to differences in nanopillar size and variations of the tunnel barriers caused by the fabrication process, the P, AP resistance levels of $A$, $B$ and $C$ differ from each other.
In addition, the average switching rate of $B$ and $C$ is higher due to lower energy barriers in these junctions.
In Figure \ref{fig:SMTJ_characterization} the characteristic RV- and PV-transfer functions are plotted.
The decline in resistance for a non-volatile MTJ state in Figure \ref{fig:SMTJ_characterization}a is attributed to the non-linear tunneling current and the resistance curve is fitted by the following approximation \cite{tunnel_current_brinkman}:
$R(V)=a/(1 + b V + c V^2) + d$, where $a, b, c$ and $d$ are fitting parameters.
\begin{figure}[t]
\centering
\includegraphics[width=0.48\textwidth]{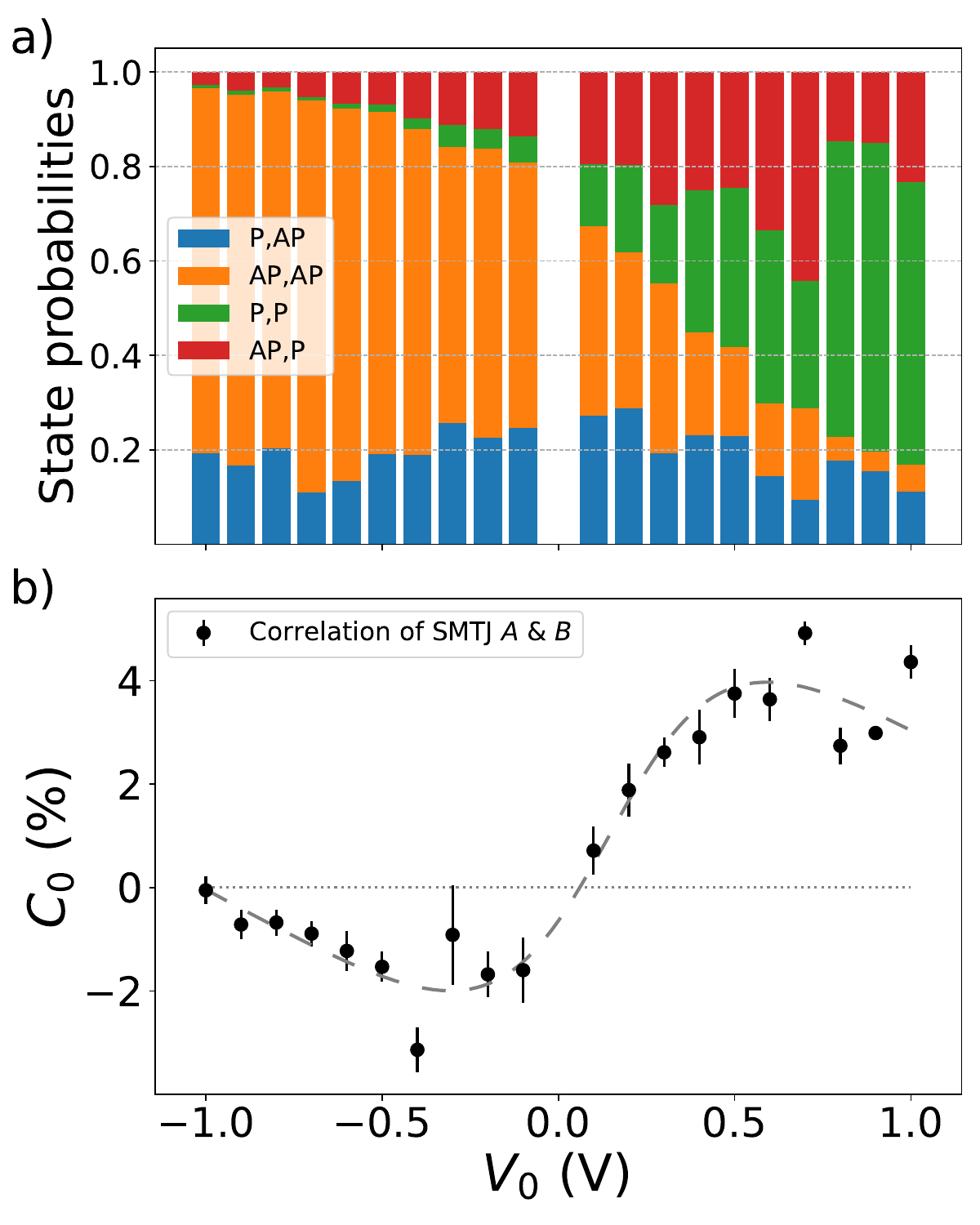}
\caption{\label{fig:state_prob_and_C0}a) The dependence of state probability on the source voltage $V_0$ for two coupled SMTJs is illustrated. b) Cross-correlation at a time lag 0 is plotted as a function of applied voltage for two coupled SMTJs. The dashed line indicates the correlation trend.}   
\end{figure}
\begin{figure}[t]
\centering
\includegraphics[width=0.48\textwidth]{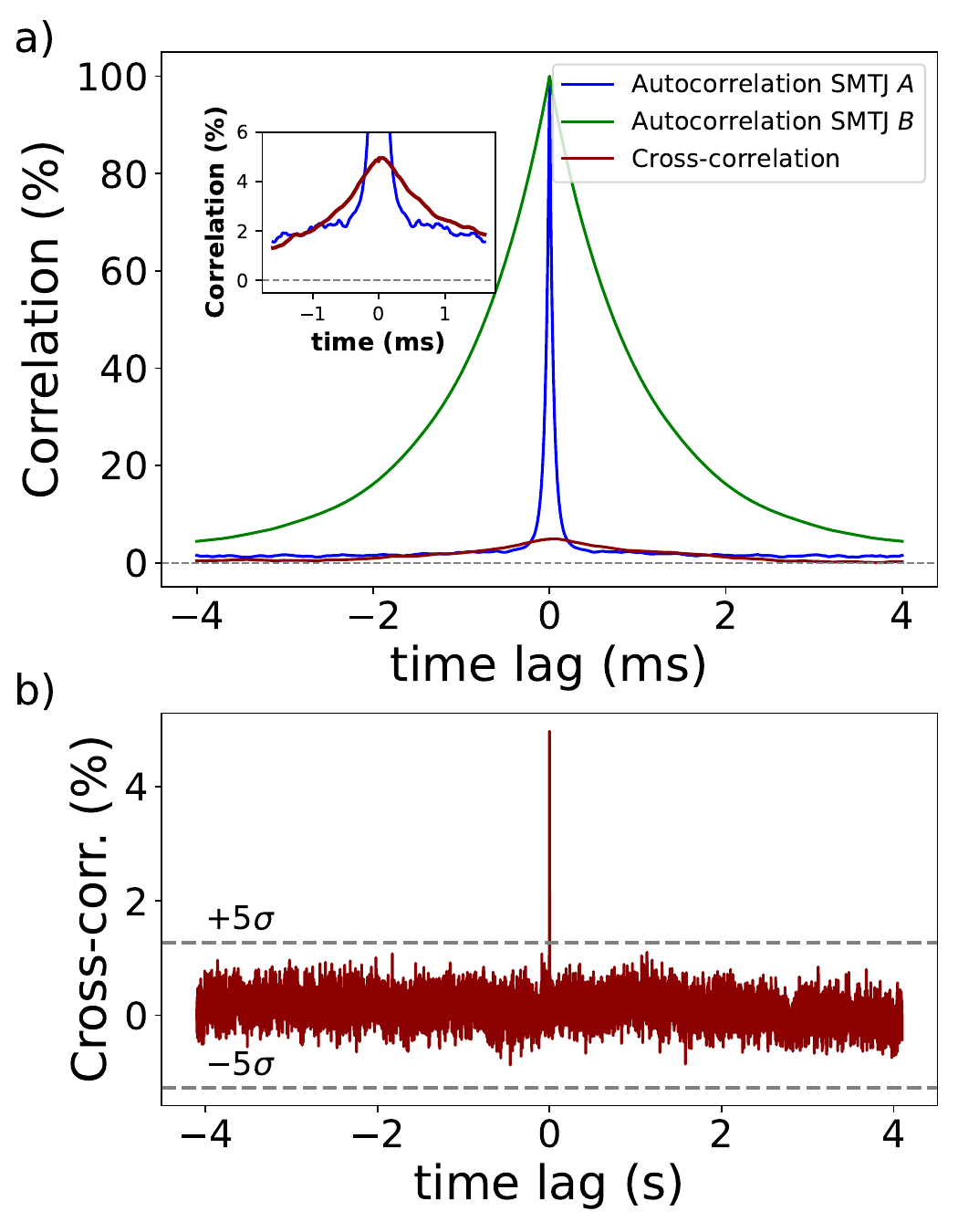}
\caption{\label{fig:autocorrelations}a) Autocorrelation functions of SMTJ $A$ and $B$ as well as their cross-correlation function are plotted ($V_0=0.7\,$V). b) Cross-correlation ($V_0=0.7\,$V) for a time lag range of $\pm4\,$s. A $\pm5\,\sigma$ level is indicated as a dashed line.}  
\end{figure}
In Figure \ref{fig:SMTJ_characterization}b each datapoint corresponds to a time series, for which the AP state probability was determined.
The applied fit is a sigmoid function based on equation \ref{eq:PV_sigmoid}.
For two SMTJs in series, there exist four different states, which also depend on the applied source voltage since each state probability is influenced by the characteristic PV-transfer curve of each SMTJ.
Figure \ref{fig:state_prob_and_C0}a highlights the state manipulation through the source voltage $V_0$, demonstrating the impact of voltage control on the coupled system.
For large negative voltages, the dominant state is the (AP,AP) state, however, in the positive voltage range, this state is less often occupied mainly for the sake of state (P,P).
This indicates that both SMTJs have the same probability trend with $V$, in which positive voltages stabilize the parallel while negative voltages stabilize the antiparallel state.
However, the coupling strength can not be easily quantified from this plot.

To investigate the coupling between both SMTJs, the cross-correlation of the time series $A$ and $B$ was analyzed.
For this, the states of SMTJ $A$ and $B$ were derived from the measured (mixed) time series, as shown in Fig. \ref{fig:setup}b.
For instance, the P state of SMTJ $A$ corresponds to the measured voltage of the coupled states of (\textbf{P},AP) \& (\textbf{P},P), whereas the AP state corresponds to all states of (\textbf{AP},AP) \& (\textbf{AP},P), as indicated in Fig. \ref{fig:setup}b.
The states of the derived binary sequence of each SMTJ are converted to a sequence of -1 and +1, representing the parallel and antiparallel states, respectively.
The normalized cross-correlation (see equation \ref{eq:xcorr}, which is equivalent to the normalized Pearson correlation coefficient with time lag $t$, is used to assess the strength of coupling.
Results are shown in Figure \ref{fig:autocorrelations}a together with the derived autocorrelations.
The cross-correlation signal exhibits a small peak of around $5\,\%$ at 0 time lag and declines exponentially for increasing time lags as outlined in the inset figure.
In Fig. \ref{fig:autocorrelations}b the same cross-correlation is illustrated over a time lag range of $\pm 4\,$s.
Here, the sharp peak at 0 indicates a significant positive coupling.
Due to the characteristic sigmoidal PV-transfer curves, the cross-correlation is expected to change with applied voltage.
Therefore, we measured the maximum correlation close to 0 time lag depending on $V_0$.
The results are shown in Figure \ref{fig:state_prob_and_C0}b.
On the positive voltage side, positive correlations can be observed, which can be attributed to the "reversed" sigmoid curve, which results in positive correlations, as elucidated in the simulation section.
Higher voltages $V_0$ lead to higher correlations due to the higher voltage drops at the junctions, thereby exerting a greater influence on the state probabilities.
Consequently, at low absolute voltages, the coupling strength diminishes, falling close to zero.
However, for high negative voltages, the correlation also approaches zero, since here almost no switching occurs because both SMTJs predominantly remain in the AP state (see Fig. \ref{fig:state_prob_and_C0}a).
Although the coupling effect persists for negative voltages, it becomes challenging to quantify it using cross-correlation techniques.
Further, the trend of correlations undergoes a sign change at $V_0=0$, changing from negative to positive correlations.
This can be explained by the "reversed" impact of negative voltages on the PV-transfer function.


In conclusion, this study highlights the successful coupling of two series-connected superparamagnetic tunnel junctions (SMTJs).
The controllability of this electrical coupling is demonstrated through the manipulation of the applied voltage.
By conducting comprehensive simulations, we establish that spin-transfer torques play a crucial role in facilitating the coupling phenomenon, which is influenced by the probability transfer function of each individual SMTJ.
The switching cross-correlation function is shown to be an effective tool for quantifying both the strength and the nature of the coupling. 
Our results reveal that SMTJs can exhibit coupling tendencies towards the same state, the opposite state, or a combination of both.
Experimental measurements confirm the existence of coupling, even in cases with different average switching rates, and indicate a preference for SMTJs to align into the same states under positive voltages.
Notably, the switching cross-correlation undergoes a sign change for negative voltages, and a peak value of approximately 5\,\% is observed at 0.7\,V.
The coupling strength can be enhanced by increasing the TMR ratio or by stronger spin-transfer torques.
Overall, we demonstrated the ability to control the stochasticity of electrically connected SMTJs, enabling the generation of a tunable probability distribution, which is a main feature in Bayesian neural networks.
In particular, the simple electrical coupling effect can be harnessed easily to enable stochastic MTJ-based Boltzmann machines.

\begin{acknowledgments}
This work was supported by the Max Planck Graduate Center with the Johannes Gutenberg-Universität Mainz (MPGC) and used infrastructure provided by ForLab MagSens.
We acknowledge support by the Deutsche Forschungsgemeinschaft (DFG, German Research Foundation) Project No. 268565370 (SFB TRR173 Projects A01 and B02), by TopDyn and the Zeiss Foundation through the Center for Emergent Algorithmic Intelligence and the Horizon Europe Project No. 101070290 (NIMFEIA) as well as ERC-2019-SyG No. 856538 (3D MAGiC). 
We would also like to thank T. Reimer for his technical support during the development of the samples
\end{acknowledgments}

\section*{Data availability}
The data that support the findings of this study are available from the corresponding author upon request.

\bibliography{bibcontent}

\end{document}